\documentclass[journal]{IEEEtran}

\usepackage{bm}
\usepackage{array}
\usepackage{threeparttable}
\usepackage{multirow}
\usepackage{booktabs}
\usepackage{graphics}
\usepackage{epstopdf}
\usepackage{epsfig} 
\usepackage{algorithm}
\usepackage{algorithmic}
\usepackage{enumerate}
\usepackage{setspace}
\usepackage{cite}
\usepackage{amsmath}
\usepackage{color}

\begin{document}

\title{Cooperative Driving at Unsignalized Intersections Using Tree Search}

\author{Huile~Xu,
        Yi~Zhang,~\IEEEmembership{Member,~IEEE,}
        Li~Li,~\IEEEmembership{Fellow,~IEEE,}
        and~Weixia~Li

\thanks{Manuscript received in February 4, 2019; This work was supported in part by National Natural Science Foundation of China under Grant 61673233, 61790565, 61603005, and the Beijing Municipal Commission of Transport Program under Grant ZC179074Z. (Corresponding author is \emph{Li Li}).}
\thanks{H. Xu is with Department of Automation, Tsinghua University, Beijing 100084, China. (E-mail: hl-xu16@mails.tsinghua.edu.cn)}
\thanks{Y. Zhang is with Department of Automation, BNRist, Tsinghua University, Beijing 100084, China and also with Berkeley Shenzhen Institute (TBSI), Tower C2, Nanshan Intelligence Park 1001, Xueyuan Blvd., Nanshan District, Shenzhen 518055, China. (E-mail: zhyi@tsinghua.edu.cn)}
\thanks{L. Li is with Department of Automation, BNRist, Tsinghua University, Beijing 100084, China. (Tel:+86(10)62782071, E-mail: li-li@tsinghua.edu.cn).}
\thanks{W. Li is with Department of Automation, Tsinghua University, Beijing 100084, China. (E-mail: wx-li11@mails.tsinghua.edu.cn)}
}

\markboth{}%
{Shell \MakeLowercase{\textit{Xu et al.}}: Cooperative Driving at Unsignalized Intersections Using Tree Search}

\maketitle

\begin{abstract}
In this paper, we propose a new cooperative driving strategy for connected and automated vehicles (CAVs) at unsignalized intersections. Based on the tree representation of the solution space for the passing order, we combine Monte Carlo tree search (MCTS) and some heuristic rules to find a nearly global-optimal passing order (leaf node) within a very short planning time. Testing results show that this new strategy can keep a good tradeoff between performance and computation flexibility.
\end{abstract}

\begin{IEEEkeywords}
Connected and Automated Vehicles (CAVs), cooperative driving, unsignalized intersection, Monte Carlo tree search (MCTS).
\end{IEEEkeywords}

\IEEEpeerreviewmaketitle

\section{Introduction}
\IEEEPARstart{C}{onnected} and Automated Vehicles (CAVs) are believed to be a key role in the next-generation transportation systems \cite{li2017recasting}. With the aid of vehicle-to-vehicle (V2V) communication, CAVs can share their driving states (position, velocity, etc.) and intentions with adjacent vehicles \cite{li2014survey,sukuvaara2009wireless} to better coordinate their motions to alleviate traffic congestion and improve traffic safety.

In the last decade, various strategies had been proposed to make optimal coordination for CAVs at a typical driving scenario: unsignalized intersection. It is pointed out in \cite{guler2014using} and \cite{li2006cooperative} that the key problem is to determine the optimal order of CAVs that passed the intersection. As summarized in \cite{meng2018analysis}, there are two kinds of cooperative driving strategies, planning based and ad hoc negotiation based, for determining the passing order.

Planning based strategies aim to enumerate all possible passing orders to find the globally optimal solution \cite{xu2018grouping}. There are two equivalent formulations of the problem. Most state-of-the-art studies formulate the problem as a mixed integer linear programming problem of vehicles' passing time scheduling \cite{li2017recasting,muller2016intersection}. The objective is usually set to minimize the total delay of all CAVs. Li \textit{et al.} showed that we can also view this problem as a tree search problem. Each tree node indicates a special (partial) passing order. The equivalent objective is to find the leaf node corresponds to the minimum total delay of all CAVs \cite{guler2014using,li2006cooperative}. It was shown in \cite{chen2016cooperative} that some planning based strategies work well for ramp metering scenarios. However, the time to enumerate all the nodes increases sharply as the number of vehicles increases, especially for unsignalized intersection scenarios. This problem hinders their applications in practice.

Ad hoc negotiation based strategies aim to find an acceptable passing order using some heuristic rules within a very short time. For example, Stone \textit{et al.} proposed autonomous intersection management (AIM) cooperative driving strategy which divides the intersection into grids (resources) and assigns these grids to CAVs in a roughly First-In-First-Out (FIFO) manner \cite{dresner2004multiagent,dresner2008multiagent}. This strategy has several variations, including reservation strategy \cite{choi2018reservation}. However, as shown in \cite{meng2018analysis}, the passing orders found by ad hoc negotiation based strategies were not good enough in many situations.

To keep a good tradeoff between performance and computation flexibility, we propose a new cooperative driving strategy based on the tree representation of the solution space for the passing order. Its key idea is to use the limited planning time to explore the nodes that are potential to be the optimal solution. To this end, we combine Monte Carlo tree search (MCTS) and some heuristic rules to accelerate the searching process, since the solution space of this problem has special structures to be exploited. Testing results show that we can find a nearly global-optimal passing order within a short enough planning time.

To give a better presentation of our finding, the rest of this paper is arranged as follows. \textit{Section II} formulates the problem and briefly reviews the existing strategies. \textit{Section III} presents the new strategy. \textit{Section IV} validates the effectiveness of the proposed strategy via numerical testing results. Finally, \textit{Section V} gives concluding remarks.

\section{Problem Formulation}
\begin{figure}[htb]
    \centering
    \includegraphics[width=7cm]{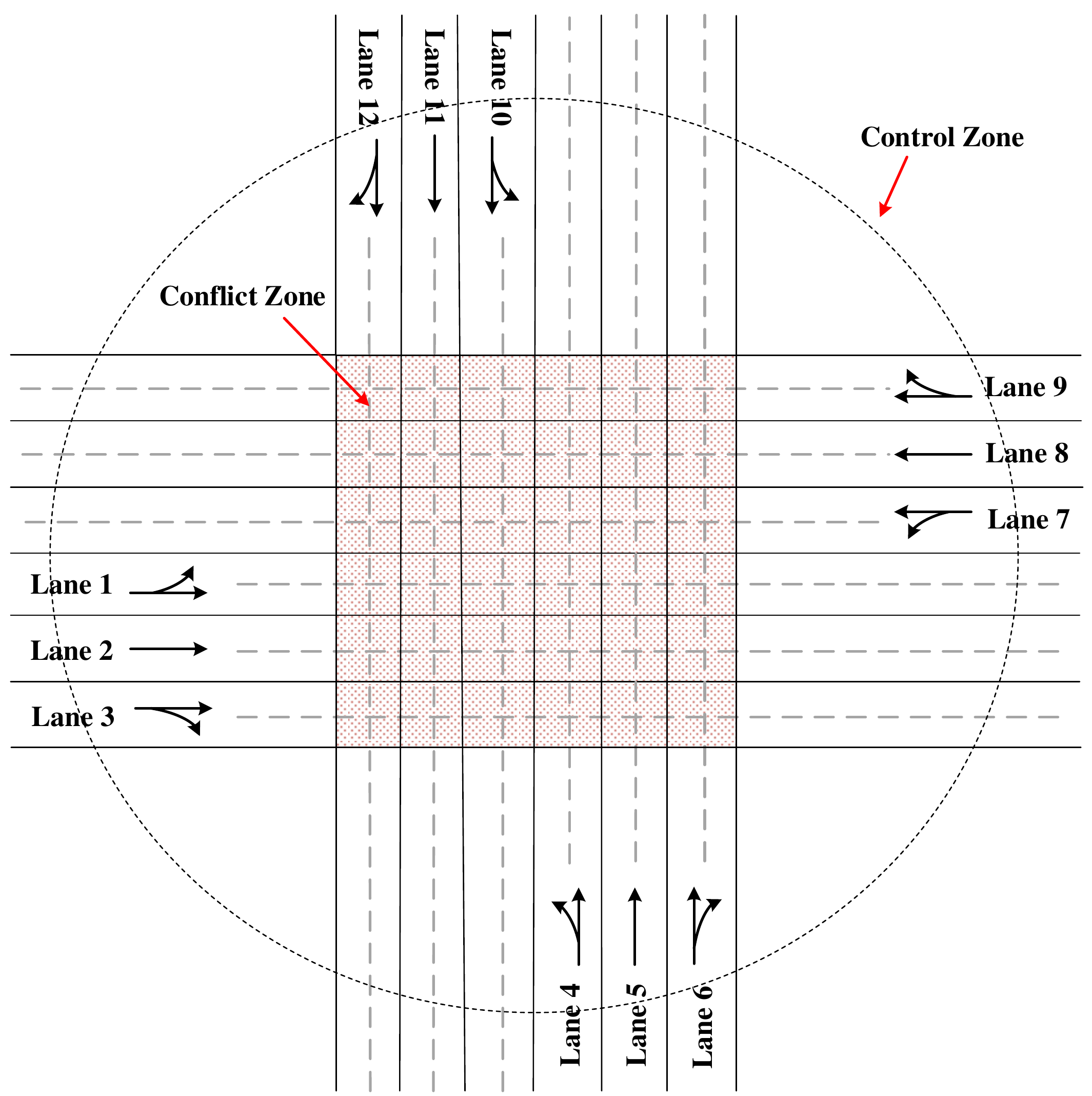}
    \caption{A typical intersection scenario.}
    \label{fig:intersection_multi_lane_scenario}
\end{figure}

Fig. \ref{fig:intersection_multi_lane_scenario} shows a typical intersection scenario with multiple lanes in each leg. The area within the circle is called the control zone, and the shadow area is called the conflict zone where lateral collisions might happen. According to the geometry of the intersection, the conflict zone can be further divided into several conflict subzones.

We assign each vehicle that enters the control zone a unique identity $V_i$.
We also use the set $Z_i$ to denote the conflict subzones that $V_i$ will pass through.
For example, $Z_i=\{4, 1\}$ means $V_i$ will pass through Conflict Subzone 4 and Conflict Subzone 1 in sequence.

To simplify the problem, we adopt the following assumptions:
\begin{itemize}
\item Each vehicle instantly and thoroughly shares its driving states (position, velocity, etc.) and intentions with other vehicles via vehicle-to-vehicle (V2V) communication.
\item Changing lane maneuver is prohibited in the control zone to ensure vehicle safety.
\item Similar to \cite{choi2018reservation} and \cite{zhang2017decentralized}, the velocities of vehicles are constant when passing through the conflict zone.
\end{itemize}

The cooperative driving strategy aims to minimize the total delay of vehicles by scheduling the velocity and acceleration profiles of all vehicles \cite{malikopoulos2018decentralized}. So, we can get the following optimization problem

\begin{equation}
\min \ J=\sum_{i=1}^n(t_{assign,i,Z_i(1)}-t_{min,i,Z_i(1)})
\end{equation}

\noindent where $t_{assign,i,z}$ is the desired arrival time to the conflict subzone $z$ for $V_i$, $t_{min,i,z}$ is the minimum arrival time to the conflict subzone $z$ when $V_i$ travels at the maximum velocity and the maximum acceleration, $Z_i(1)$ is the first element in the set $Z_i$, $n$ is the number of vehicles in the control zone.

To directly attack Problem (1) often leads to a mixed integer linear programming (MILP) problem whose computation time increases exponentially with the increase of the number of vehicles \cite{muller2016intersection,chen2016cooperative}.

Noticing that the traffic efficiency mainly depends on the passing order of vehicles \cite{meng2018analysis}, we can formulate the whole problem as a tree search problem in the solution space that consists of all possible passing orders. Each leaf node represents a passing order of vehicles which can also be denoted as a string \cite{li2006cooperative}. For example, string CAB means vehicle C, vehicle A, and vehicle B enter the conflict zone sequentially.

\begin{figure}[htb]
    \centering
    \includegraphics[width=6cm]{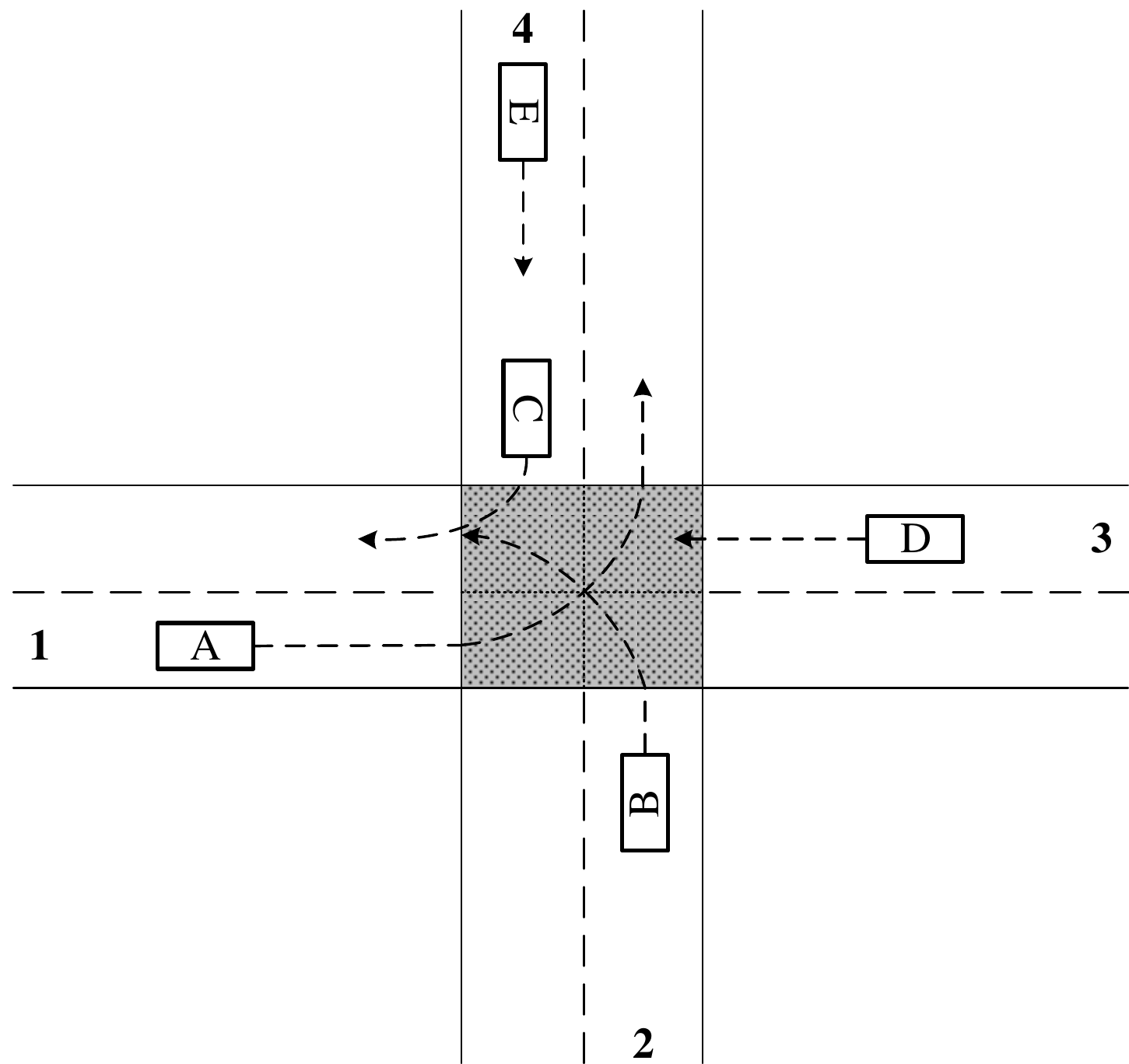}
    \caption{An intersection scenario with 5 vehicles.}
    \label{fig:intersection}
\end{figure}

\begin{figure}[htb]
    \centering
    \includegraphics[width=6cm]{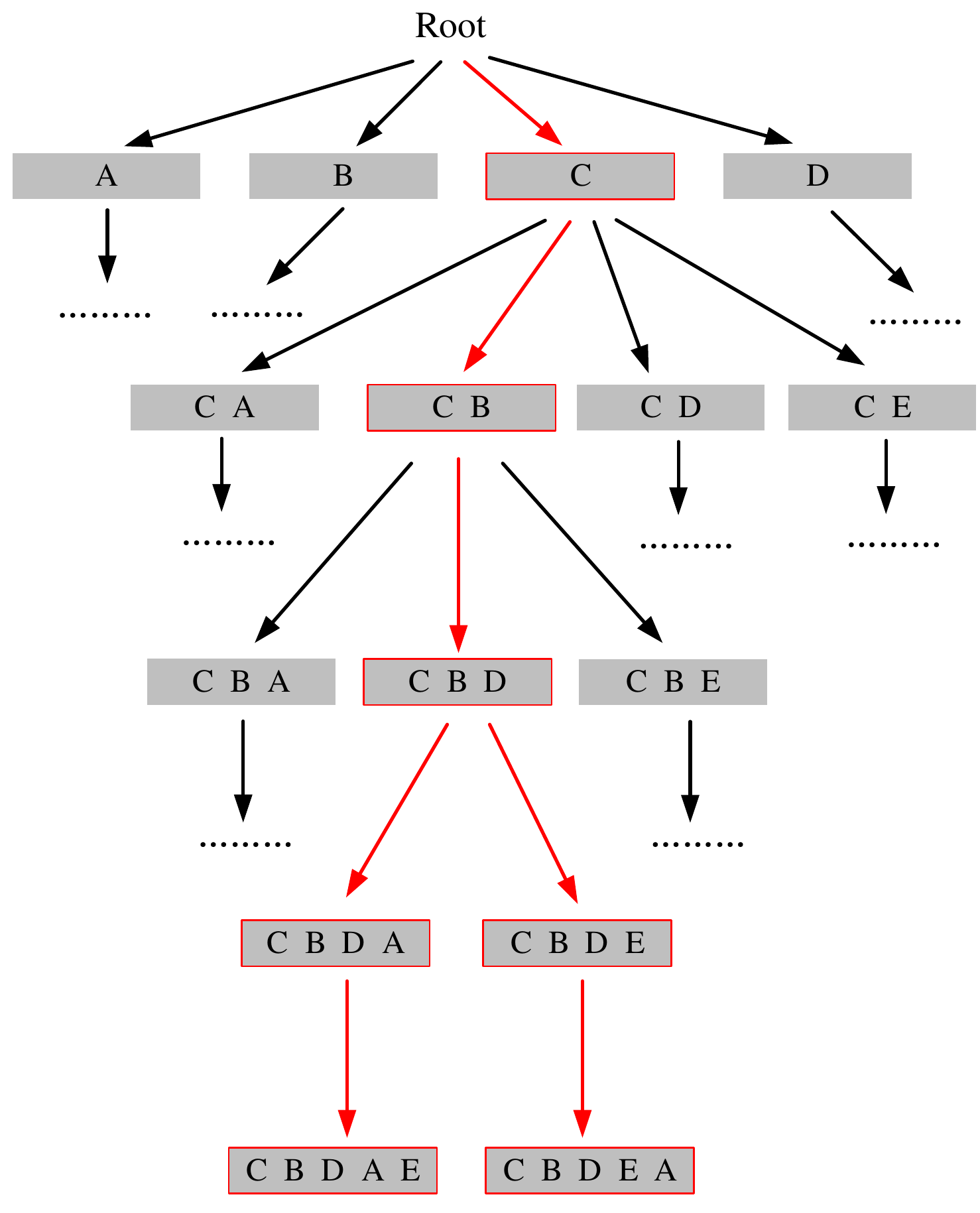}
    \caption{The solution tree stemmed from the intersection scenario shown in Fig. 2. The leaf nodes in the bottom layer represent the complete passing orders for all vehicles.}
    \label{fig:solution_tree}
\end{figure}

Let us take the intersection scenario shown in Fig. \ref{fig:intersection} as an example to explain how to build the tree representation of the solution space gradually . At first, we set the passing order in the root node to be empty. Then, each direct child node of the root node (in the second layer) refers to one index symbol that indicates the first vehicle in a special passing order. The nodes in the third layer refer to one string consisting of two indices symbols that indicate the first two vehicles in a special passing order. Similarly, the child nodes expand their child nodes, and all possible passing order are generated as leaf nodes in the bottom layer of the solution tree as shown in Fig. \ref{fig:solution_tree}.

If a (partial) passing order is given, the desired arrival times for all the vehicles that has been covered in this (partial) passing order can be directly derived by the following Passing Order to Trajectory Interpretation Algorithm.
Our objective turns to seek the leaf node that corresponds to the shortest total delay.
Moreover, the total delay values of leaf nodes can be used to evaluate the potential of their parent nodes in a backpropagation way.
This method provides us a chance to find a nearly global-optimal leaf node but only search a small part of the whole tree.

\begin{algorithm}
\caption{Passing Order to Trajectory Interpretation}
\begin{algorithmic}[1]
\begin{spacing}{1.2}
\REQUIRE A (partial) passing order $P$
\ENSURE The total delay $J$ of the covered vehicles and their arrival times $t_{assign}$
\FOR {each $i \in [1,length(P)]$}
\FOR {each $z \in Z_i$}
\STATE $V_j$ is the last vehicle that passed through subzone z
\STATE
$t_{assign,P(i),z}=max(t_{min,P(i),z}, t_{max,z}+\Delta_{j,a})$
\ENDFOR
\STATE
Adjust $t_{assign,P(i),z}$ according to the constraint: the velocity of $V_i$ in the conflict zone is constant.
\FOR {each $z \in Z_i$}
\STATE $t_{max,z}=t_{assign,P(i),z}$
\ENDFOR
\ENDFOR
\STATE $J=\sum_{i=1}^{length(P)}t_{assign,P(i),Z_i(1)}$
\end{spacing}
\end{algorithmic}
\end{algorithm}

In \textbf{Algorithm 1}, $P(i)$ is the $i$th element in the input (partial) passing order, $t_{max,z}$ is the largest arrival time that the subzone $z$ has been occupied. $\Delta_{j,a}$ is the minimum safety gap between two consecutive vehicles passing through the same subzone. Obviously, the time complexity of \textbf{Algorithm 1} is $O(n)$.
A detailed explanation of Algorithm 1 can be found in our previous report \cite{xu2018grouping}.

\section{MCTS Based Cooperative Driving Strategies}

It is usually impossible to expand all the nodes of the solution tree within the limited computation budget, when there are lots of vehicles in the control zone. In this paper, we use MCTS + heuristic rules to select nodes with the potential to be the optimal solution. The recent success of the MCTS method in the game of Go shows it is an effective way to deal with such problems \cite{enzenberger2010fuego,silver2017mastering}.

\begin{figure*}[htb]
    \centering
    \includegraphics[width=12cm]{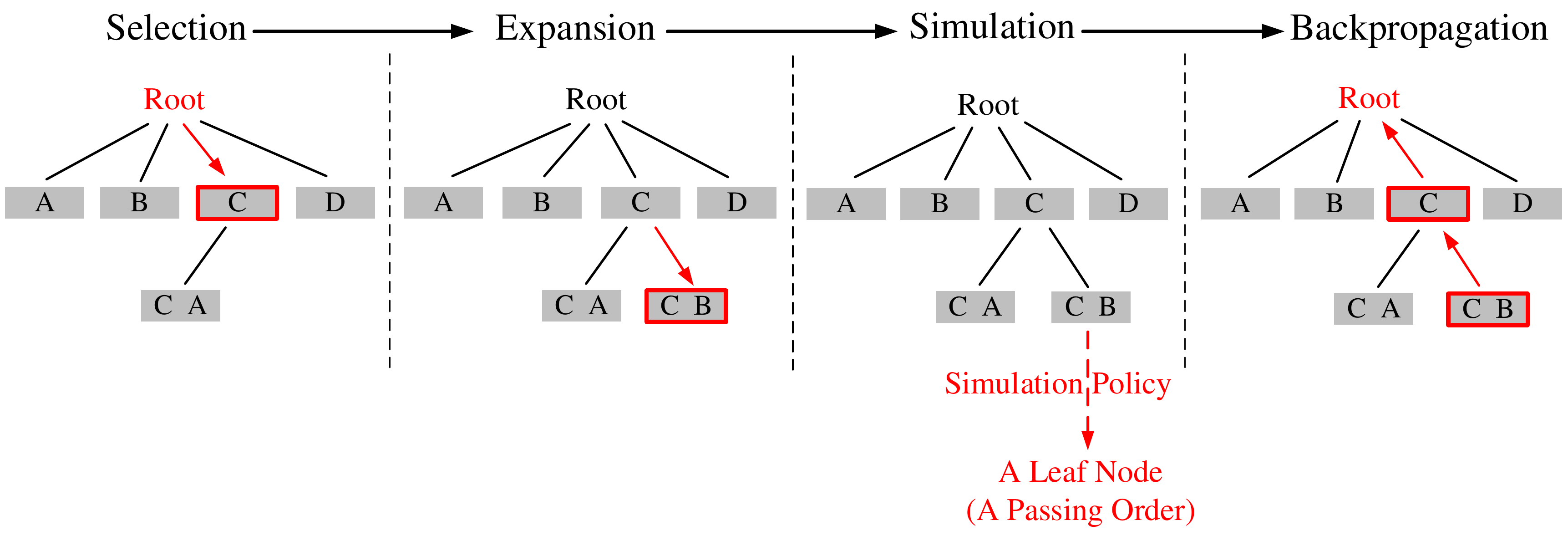}
    \caption{One iteration of the MCTS based cooperative driving strategy.}
    \label{fig:monte_carlo_tree}
\end{figure*}

\subsection{The Classical MCTS Based Strategy}

In MCTS, each node in the formulated tree will be assigned a score to evaluate its potential.
The score of a leaf node is equal to the total delay of its corresponding passing order.
MCTS uses these scores to determine which branch of tree to explore.

Generally, MCTS gradually builds a search tree in an iteration way.
One iteration consists of four steps: selection, expansion, simulation, and backpropagation \cite{browne2012survey}; see Fig. \ref{fig:monte_carlo_tree}.

\begin{enumerate}
\item Selection: Starting at the root node, we select the most urgent expandable node based on the following policy \cite{kocsis2006bandit}
    \begin{equation}
    \mathop{\arg\max}_{i} \ Q_i+C\sqrt{\frac{\ln n}{n_i}}
    \end{equation}
    \noindent where $Q_i$ is the score of child node $i$ and the value of $Q_i$ is within $[0,1]$. $n$ is the number of times the current node has been visited, $n_i$ is the number of times child node $i$ has been visited, and $C$ is a weighting parameter. The child node with the largest total score is selected. Here, an expandable node refers to a node that is not a leaf node and has unvisited child nodes.

    This child node selection policy is suggested in the field of computer Go and is called UCB1 \cite{kocsis2006bandit}. The first term in the equation encourages to select the child node that is currently believed to be optimal, while the second term encourages to explore more child nodes.
\item Expansion: We randomly select one unvisited child node of the most urgent expandable node to be a new node that is added to the tree.
\item Simulation: We run several rollout simulations to determine a complete passing order based on the partial passing order represented by the current new node to evaluate the potential of the new node.

     The classical MCTS randomly samples and adds the uncovered vehicles into the passing order string one by one, until we find a complete passing order string and reach the maximum depth of the tree from the current new node without branching \cite{silver2017mastering}. For example, when we apply random sampling policy to the node CB shown in Fig. \ref{fig:monte_carlo_tree}, we can randomly expand a direct child node in its next layer; say node CBA. The node CBA will be further expanded by repeating such a process until a leaf node (e.g., node CBADE) is reached. Finally, the partial passing order will be evaluated by all its simulated off-spring leaf nodes (passing orders). Sometimes, the generated passing order is not invalid, because it may violate the prohibition of lane change, such solutions will be discarded after check.

    After simulation, we update the scores of the current new node as follows:
\begin{enumerate}[i]
  \item Apply \textbf{Algorithm 1} to calculate the total delay $\bar{J}_{i}$ of the partial order corresponds with the current new node.
  \item Apply \textbf{Algorithm 1} to calculate the total delay $\hat{J}_{i}$ of the partial order corresponds with the best off-spring node of the current new node via simulation.
  \item Calculate the score $Q_i$ of the current new node as
    \begin{equation}
    Q_i=\omega \bar{J}_{i}+(1-\omega) \hat{J}_{i}
    \end{equation}
    where $\omega$ is a weighting parameter. Since $Q_i \in [0,1]$, we normalize $\bar{J}_{i}$ and $\hat{J}_{i}$ into $[0,1]$ before updating $Q_i$.
\end{enumerate}

\item Backpropagation: The simulation result is backpropagated through the selected nodes to update the scores of all its parent nodes.
\end{enumerate}

During the building process of the search tree, the state-of-the-art best passing order is continuously updated. As soon as the computation budget is reached, the search terminates and returns the state-of-the-art best passing order. The planned arrival times of vehicles can be determined by using \textbf{Algorithm 1}. The velocity and acceleration profiles of each vehicle plan will be finally calculated by using the motion planning method proposed in \cite{malikopoulos2018decentralized}.

We can see that the performance of the proposed strategy is influenced by the choice of the parameters including the maximum search time and two weighting parameters $C$ and $\omega$. We will discuss how to choose these parameters in \textit{Section IV} below.

\subsection{The MCTS + Heuristic Rules}

As aforementioned, the classical MCTS strategy uses random sampling to generate a leaf node (a passing order) in the simulation step. However, because of the huge number of possible passing orders, the passing orders generated by random sampling cannot help us quickly capture the real potential of a node during simulation.

Thus, we propose the following heuristic rules to help decide which nodes (vehicles) should be expanded (added into the candidate passing order string) during simulation.
Heuristic rule 1 helps to quickly prune the invalid passing order \cite{li2006cooperative}.
Heuristic rule 2 determines the vehicle among the candidates to be chosen.

\begin{enumerate}
\item For the vehicles on the same lane, the vehicle which is the closest to the conflict zone should be added earlier than other vehicles since changing lane maneuver is prohibited.
\item For the vehicles passing through the same conflict subzone, the vehicle with a less desired arrival time should be added earlier.
\end{enumerate}

The simulation step can be summarized as \textbf{Algorithm~2}.
We can see that the classic MCTS applies random sampling in both expansion and simulation steps; while our MCTS + heuristic rules applies random sampling only in expansion step.

The Ad hoc negotiation based strategies organize all the vehicles according to the FIFO principle. In contrast, the new simulation policy tends to organize just a part of vehicles (the vehicles uncovered in the current partial passing order) according to the FIFO strategy.
This trick helps to avoid the convergence to a over-greedy solution.

\begin{algorithm}
\caption{Heuristic Simulation Policy}
\begin{algorithmic}[1]
\REQUIRE Locations and velocities of all vehicles
\ENSURE A possible passing order
\STATE Among all uncovered vehicles, we select the vehicles which are the closest to the conflict zone in each lane as candidate vehicles and calculate their arrival times to all conflict subzones.
\STATE If there exists a candidate vehicle whose arrival times to all conflict subzones are all the smallest, we add it into the passing order string. If not, we randomly select one vehicle among all candidate vehicles and add it into the passing order string.
\STATE Then we repeat the steps 1 and 2 until a complete passing order string is generated.
\STATE The objective value (1) of the generated passing order can be easily derived by \textbf{Algorithm 1} and denoted as $q_{2,i}$.
\end{algorithmic}
\end{algorithm}

\section{Simulation Results}

\subsection{Simulation Settings}
We design three experiments to determine the best parameter set for the new cooperative driving strategy and compare it with some classical ones. These experiments are conducted for the intersection with three lanes in each leg shown in Fig. \ref{fig:intersection_multi_lane_scenario}. The mandatory signs stipulate the permitted directions for each lane. According to the geometry of the intersection, the conflict zone is further divided into 36 subzones. The vehicles¡¯ arrival is assumed to be a Poisson process. We vary the mean value of this Poisson process to test the performance of the proposed strategy under different traffic demands. The vehicles¡¯ arrival rates at all lanes are the same unless otherwise specified. It should be pointed out that we had tested other intersections with different road geometries and various vehicle arrival patterns, but the conclusions remain unchanged.

To accurately describe the total delays of vehicles, we adopt the point-queue model in the simulation \cite{meng2018analysis,ban2012continuous}. The model assumes vehicles travel in free flow state until it gets to the boundary of the intersection we study. If the preceding vehicle leaves enough spaces, the first vehicle in the point-queue will dequeue and enter the intersection. Otherwise, it will stay in the virtual queue. Each lane has an independent point-queue.

In this paper, we reschedule the passing order of all the vehicles within the control zone every 2 seconds.
As suggested in \cite{meng2018analysis}, we set the minimum safety gap between two consecutive vehicles passing through the same subzone as a slightly enlarged constant as
\begin{equation}
\Delta_{j,a}=
\begin{cases}
1.5 s & a=1~~~(\text{go straight})\\
2 s & a=2~~~(\text{turn left})\\
1.5 s & a=3~~~(\text{turn right})
\end{cases}
\end{equation}
to avoid the collisions caused by position measurement errors and communication delay.

\subsection{The Choice of Parameters}
In this paper, we consider two performance indices: the delay $J$ of the given $n$ vehicles and the traffic throughput (the number of vehicles that has passed the intersection control zone) within a given time interval to compare different cooperative driving strategies.
Specially, we highlight the decreased ratio of the total delay if being compared withe baseline solution that is gotten by the FIFO strategy
\begin{equation}
\eta=\frac{J_{FIFO}-J_{MCTS}}{J_{FIFO}}
\end{equation}
\noindent where $J_{FIFO}$ is the objective value of the FIFO passing order, and $J_{MCTS}$ is the objective value of the best passing order from the MCTS based strategy.

To determine the best parameter setting of the new MCTS + heuristic rules, we first fix the time budget as 0.1 s and vary $\omega$ and $C$ from 0 to 1.
To better understand the performance of the strategy under different traffic conditions, we vary the vehicle arrival rate to generate a series of intersection scenarios with different number of vehicles.

\begin{figure}[htb]
    \centering
    \includegraphics[width=8cm]{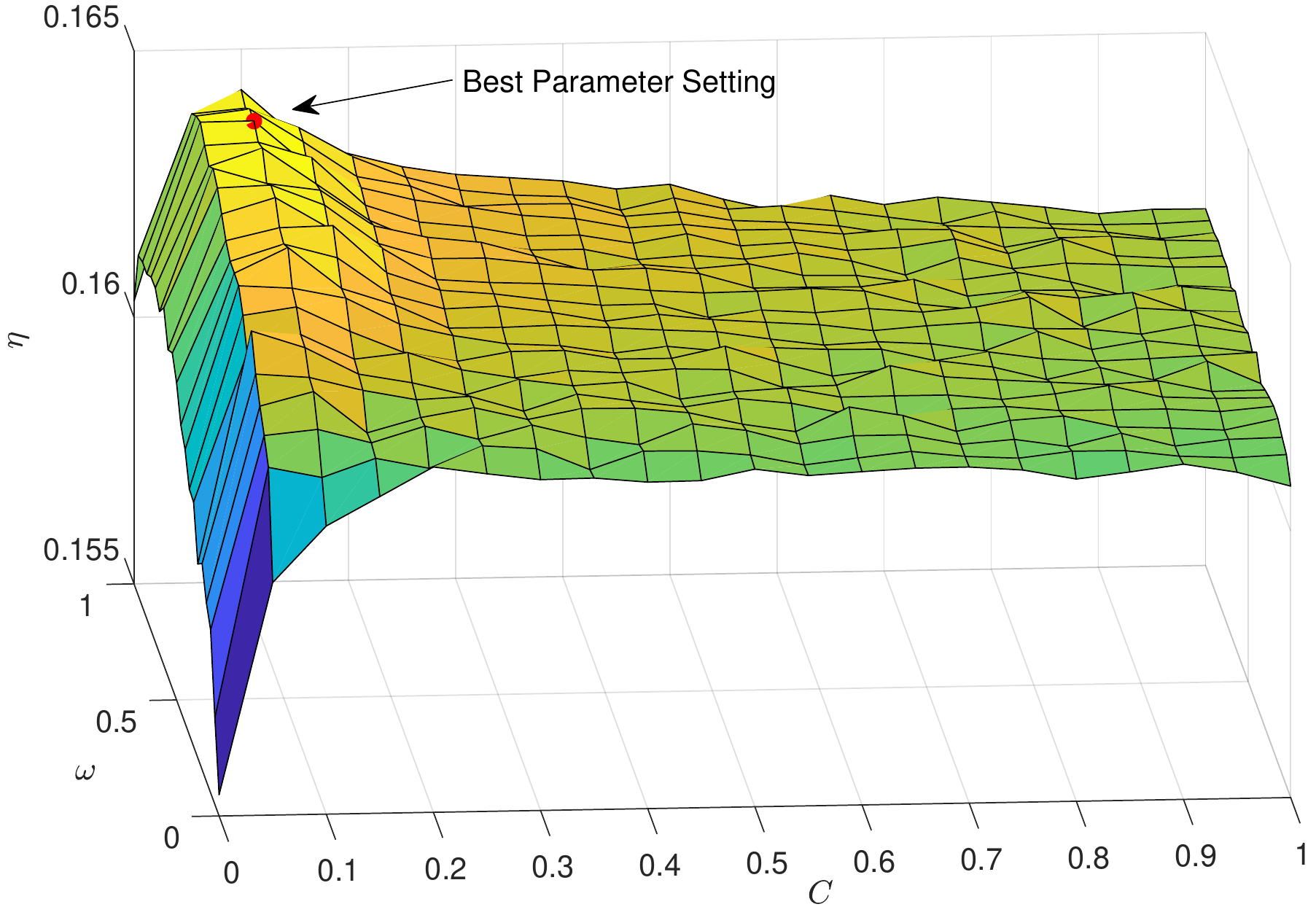}
    \caption{The improvement rate of the MCTS based strategy with different parameter settings for the intersection shown in Fig.1.}
    \label{fig:weightParameters30}
\end{figure}

Fig. \ref{fig:weightParameters30} gives the improvement rates for the intersection shown in Fig.1 with 30 vehicles.
We can see that a significant improvement can be achieved even with the worst parameter setting.
The parameter $C$ and $\omega$ are not so critical but may still influence the balance between exploitation and exploration, partly because we use heuristic rules in simulation step to reduce the influence of random sampling.
We further study the scenarios with other numbers of vehicles and the results are all similar.
Thus, in the rest of this paper, we set $\omega = 0.85$ and $C = 0.05$.

Then, to determine an appropriate time budget, we vary the time limits of tree search.
To eliminate the influence of the computing power of the device, we examine the improvement rates with respect to the number of nodes that has been searched.

\begin{figure}[htb]
    \centering
    \includegraphics[width=8cm]{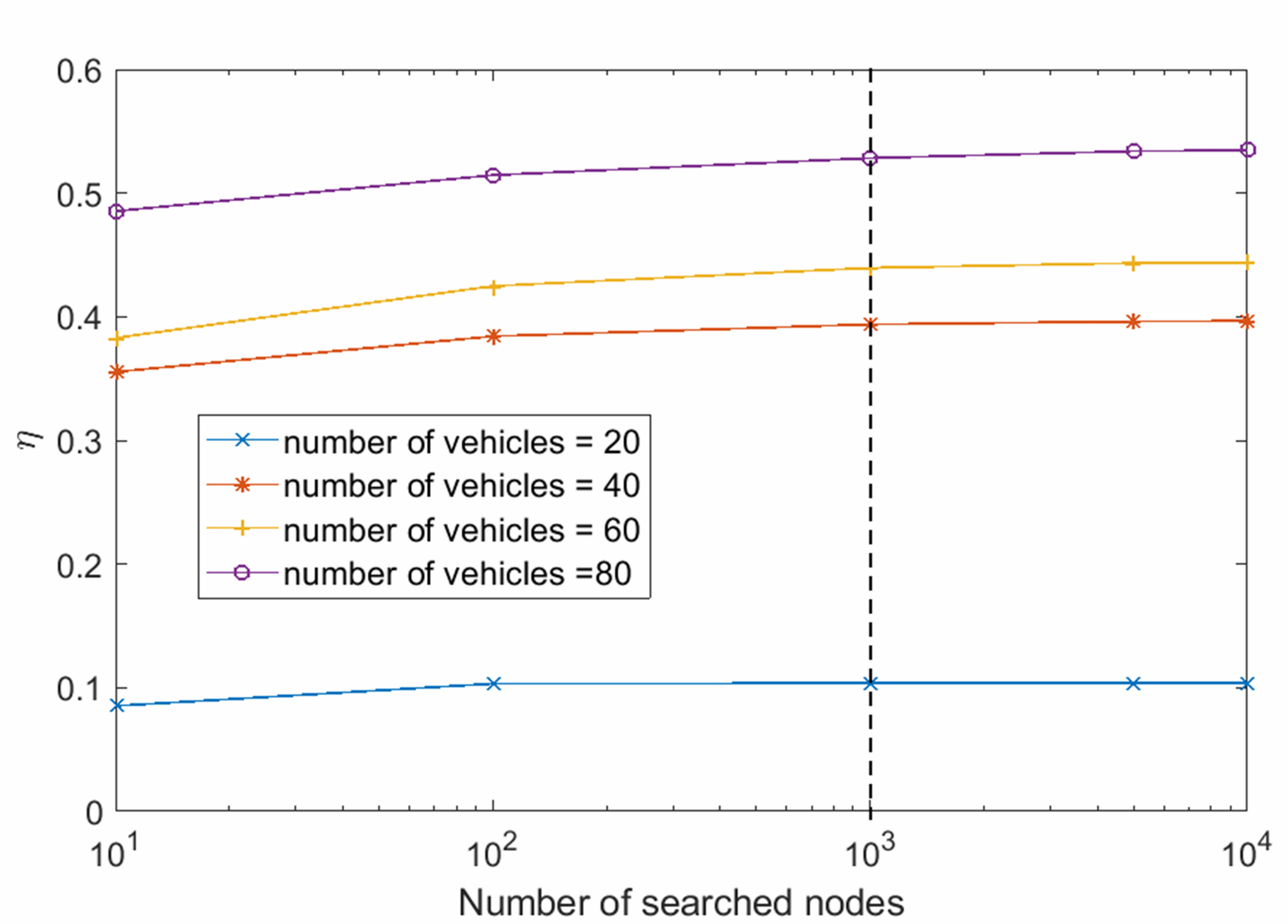}
    \caption{The results of the improvement rates with respect to the number of searched nodes for the intersection shown in Fig.1.}
    \label{fig:nodes}
\end{figure}

It can be seen from Fig. \ref{fig:nodes} that the improvement rate increases significantly when the number of searched nodes increases from 10 to 1000.
However, the improvement rate soon becomes saturated after that.
Thus, we believe that the proposed strategy can obtain a good enough passing order through searching 1000 nodes.
For most intersection scenarios, 1000 nodes can be searched within 0.1 s in our personal computer, so we set the maximum search time as 0.1 s for the following experiments.

\subsection{Comparisons of Different Cooperative Driving Strategies}

To further clarify the difference between the FIFO strategy and our new strategy, we study a typical intersection scenario with single lane in each leg and 20 vehicles.
We calculate the objective values for all the valid solutions (passing orders) and plot them in a histogram manner; see Fig. \ref{fig:intersection_multi_lane_scenario}.

\begin{figure}[htb]
    \centering
    \includegraphics[width=8cm]{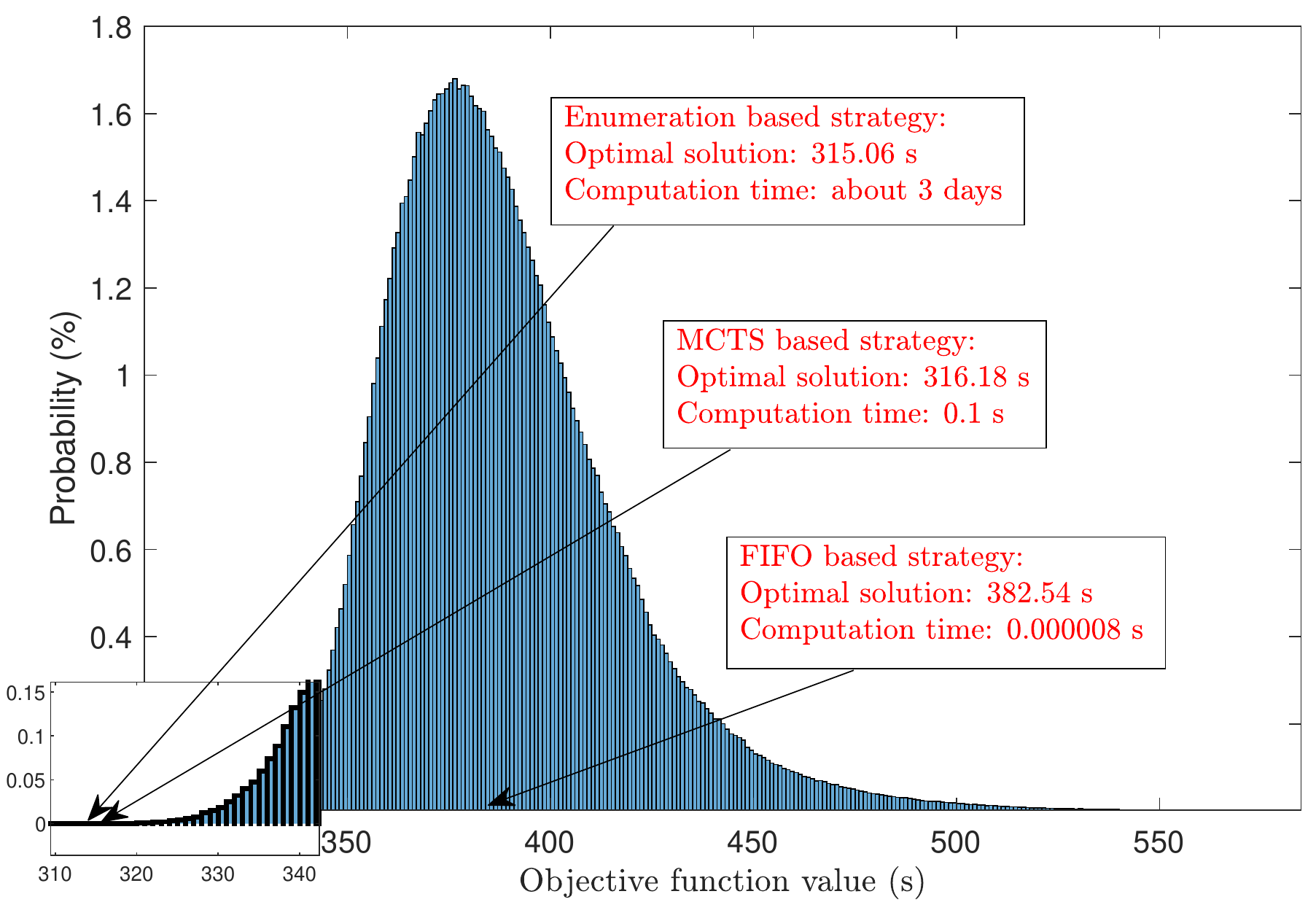}
    \caption{The histogram of all solution values for a single lane intersection scenario with 20 vehicles.}
    \label{fig:histogram}
\end{figure}

It is clear that the solution found by the MCTS based strategy is nearly the same as the global optimal solution found by the enumeration based strategy, while the computation time of the MCTS based strategy is much less. For the FIFO based strategy, the computation time is the least, but the solution is far away from the optimal solution. The solution found by the MCTS based strategy ranks 648th in the nearly 10 billion solutions; while the solution of the FIFO based strategy ranks 4563421793th.

We then carry out another comparison for the intersection shown in Fig. \ref{fig:intersection_multi_lane_scenario}, where the average arrival rate is varied to explore the influence of different traffic demands.
For each arrival rate, we simulate 20-minute.
It is obvious that our new strategy further reduces the average delay and improves the traffic throughput in all situations.

\begin{table}[htbp]
\footnotesize
  \centering
  \caption{Comparison results of different cooperative driving strategies}
    \begin{tabular}{cp{4.235em}cc}
    \toprule
    \multicolumn{1}{p{5.2em}<{\centering}}{Arrival rate veh/(lane*h)} & Strategies & \multicolumn{1}{p{6.5em}<{\centering}}{Average delay (s)} & \multicolumn{1}{p{5.2em}<{\centering}}{Traffic Throughput (veh)} \\
    \midrule
    \multirow{2}[2]{*}{150} & FIFO  & 1.3053 & 589 \\
          & MCTS  & 0.4499 & 605 \\
    \midrule
    \multirow{2}[2]{*}{300} & FIFO  & 39.8313 & 1095 \\
          & MCTS  & 1.1407 & 1168 \\
    \midrule
    \multirow{2}[2]{*}{450} & FIFO  & 41.6996 & 1205 \\
          & MCTS  & 4.8743 & 1766 \\
    \bottomrule
    \end{tabular}%
    \begin{tablenotes}
            \footnotesize
            \item{$^*$} The computation time of the MCTS based strategy is 0.1s.
    \end{tablenotes}
  \label{tab:tab1}%
\end{table}%

\subsection{A Further Look into the Structure of the Obtained Search Tree}

\begin{figure}[htb]
    \centering
    \includegraphics[width=7cm]{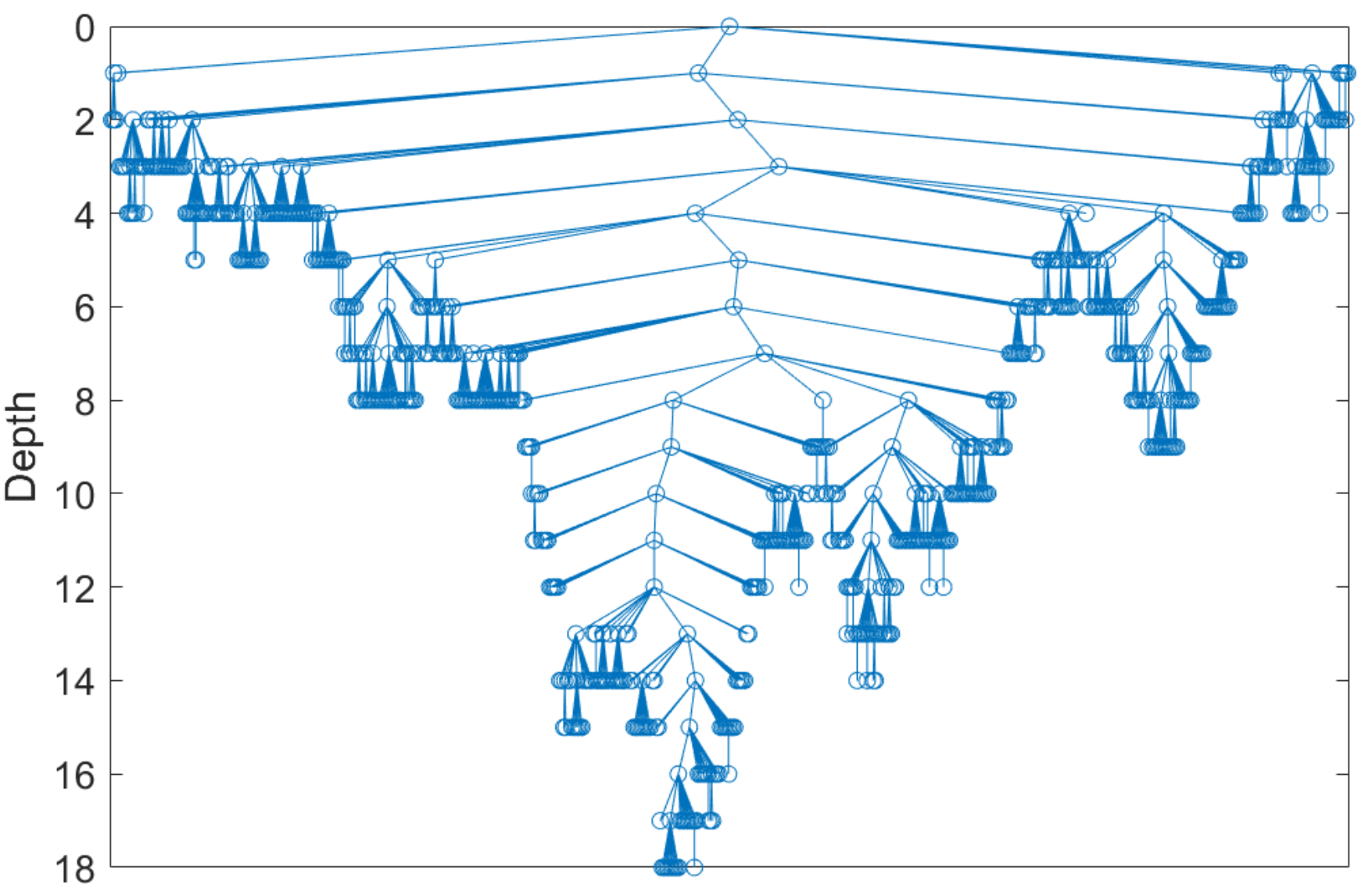}
    \caption{The structure of a search tree for an intersection scenario with 50 vehicles.}
    \label{fig:treePlot50}
\end{figure}

Fig. \ref{fig:treePlot50} shows the formulated search tree of our new strategy for an intersection scenario with 50 vehicles.
Similar to the classical MCTS strategy, our new strategy tends to first find some promising branches (partial passing orders) of the tree and spends most search time to further explore these branches.
However, the search tree generated by the classical MCTS strategy contains much more unnecessary leaf nodes.
In contrast, when the heuristic rules are introduced, only a very small number of leaf nodes will be finally reached.
For this case, although there are more than $10^{46}$ possible passing orders, only about two thousands passing orders are explored by our new strategy within 0.1 s.
This difference explains why the classical MCTS needs much more time to find a good enough passing order.

\section{Conclusion}
In this paper, we propose a cooperative driving strategy that combines Monte Carlo simulation and heuristic rule simulation to accelerate the search of the passing order.
This new method can quickly learn the tree structure knowledge of the given scenario and find a nearly optimal solution with a short time.
Although we only discuss the schedule of vehicles at unsignalized intersections, this method can be easily adapted to other scenarios (e.g., ramping areas and working zones).
We are currently building several automated vehicle prototypes so that we can test our new strategy in field studies in the near future.

\end{document}